\documentclass[a4paper]{jpconf}
\usepackage{graphicx}
\usepackage{amssymb}
\usepackage{color,hyperref}
\definecolor{darkblue}{rgb}{0.0,0.0,0.3}
\hypersetup{colorlinks,breaklinks,
            linkcolor=darkblue,urlcolor=darkblue,
            anchorcolor=darkblue,citecolor=darkblue}
\usepackage[square,sort&compress,numbers]{natbib}
\newcommand\apj{\rmfamily{ApJ}}%
\newcommand\apjl{\rmfamily{ApJ}}%
\newcommand\apss{\rmfamily{Ap\&SS}}%
\newcommand\aaps{\rmfamily{A\&AS}}%
\newcommand\mnras{\rmfamily{MNRAS}}%
\newcommand\na{\rmfamily{New Ast.}}%
\newcommand\prd{\rmfamily{Phys.~Rev.~D}}%
\newcommand\pasj{\rmfamily{PASJ}}%
\newcommand\nat{\rmfamily{Nature}}%
\newcommand{\OmegaH}{\Omega_\mathrm{H}}

\newcommand{\avg}[1]{\ensuremath{\langle#1\rangle}} 
\newcommand{\abs}[1]{\ensuremath{\left|#1\right|}}

\newcommand{\MdotH}{\dot M}

\newcommand{\etaj}{\eta_{\rm jet}}
\newcommand{\etaw}{\eta_{\rm wind}}

\newcommand{\etabz}{\eta_{\rm BZ}}

\newcommand{\phibh}{\phi_{\rm BH}}
\newcommand{\PhiBH}{\Phi_{\rm BH}}

\newcommand{\cut}[1]{\hbox{}}

\begin{document}
\pagestyle{plain}

\title{General Relativistic Modeling of Magnetized Jets from Accreting
  Black Holes}

\author{Alexander Tchekhovskoy$^1$, Jonathan C. McKinney$^2$ and
  Ramesh Narayan$^3$}

\address{$^1$Center for Theoretical Science, Jadwin Hall, Princeton University, Princeton,
  NJ 08544, USA; Princeton Center for Theoretical Science Fellow}
\address{$^2$Kavli Institute for Particle Astrophysics and Cosmology, Stanford University, P.O. Box 20450, MS 29,
Stanford, CA 94309, USA}
\address{$^3$Institute for  Theory and Computation, Harvard-Smithsonian Center for Astrophysics,
 60 Garden Street, MS 51, Cambridge, MA 02138, USA}

\ead{atchekho@princeton.edu}

\begin{abstract}
  Recent advances in general relativistic magnetohydrodynamic modeling of jets offer
  unprecedented insights into the inner workings of accreting black
  holes that power the jets in active galactic nuclei (AGN) and other
  accretion systems. I will present the results of recent
  studies that determine spin-dependence of jet power and discuss the
  implications for the AGN radio loud/quiet dichotomy and recent
  observations of high jet power in a number of AGN.
\end{abstract}

\section{Introduction}
Relativistic jets are one of the most spectacular
manifestations of black hole (BH) accretion.  Jet-producing accretion
systems span $9$ decades in central BH mass: from stellar-mass
BHs
in black hole binaries (BHBs) and
gamma-ray bursts (GRBs) to supermassive
BHs 
in active galactic nuclei (AGN). If a
single physical mechanism is responsible for producing jets throughout
the BH mass spectrum, it must be robust and scale
invariant. Magnetic fields are a promising agent for jet production because they
are abundant in astrophysical plasmas and because the properties of magnetically-powered
jets scale trivially with BH mass
\citep{bz77,chiueh_asymptotic_jet_structure_91,hs03,tch08,tch09}.

\begin{figure}[t]
\begin{center}
\includegraphics[width=0.8\textwidth]{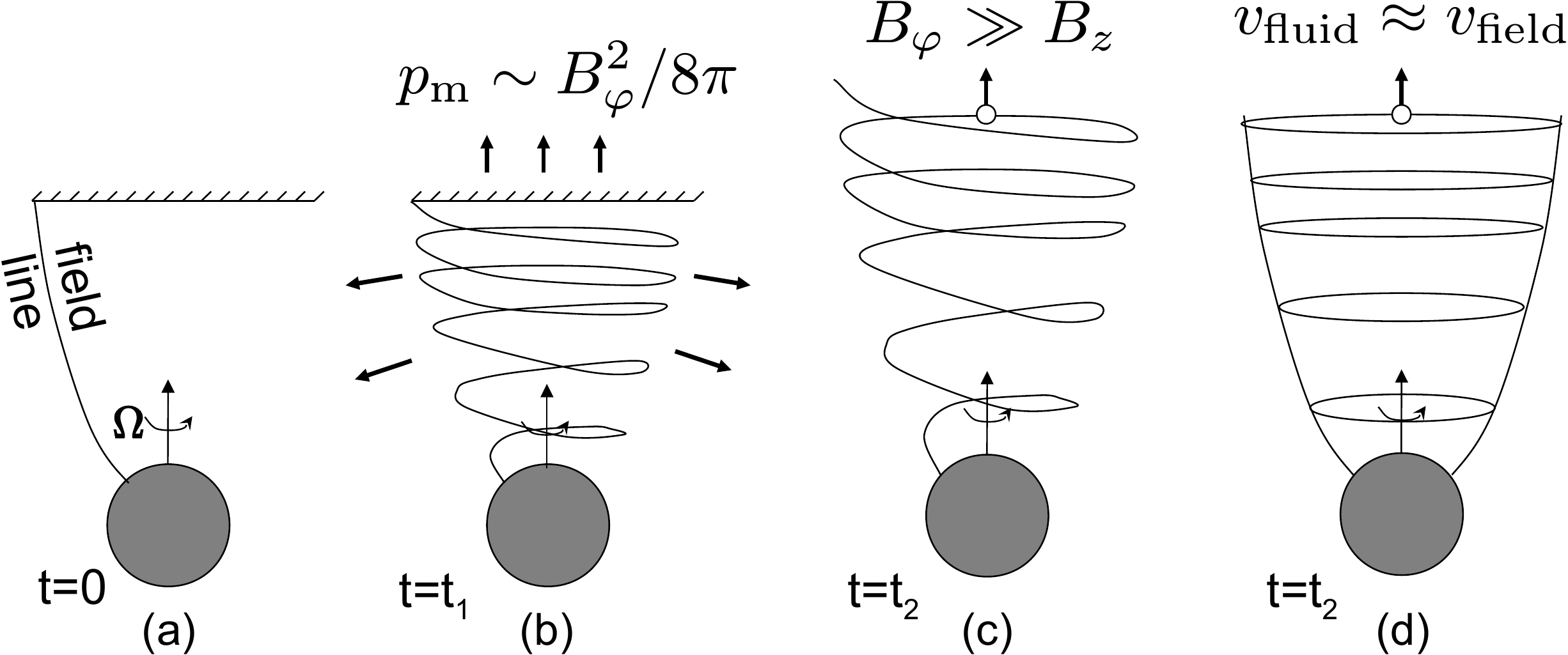}
\end{center}
\vspace{-0.5cm}
\caption{\label{jetcartoon} Illustration of jet formation by magnetic fields.
(a)~Consider a purely poloidal (i.e., $B_\varphi=0$)
field line attached on one end to a stationary ``ceiling'' (which
represents the ambient medium and is shown with hashed horizontal
line) and on the other end to a
perfectly conducting sphere (which represents the
central BH or disk and is shown with grey filled circle)
rotating
at an angular frequency~$\Omega$.  (b)~After $N$
rotations, at time $t=t_1$, the initially purely poloidal field line
develops $N$ toroidal loops.  This magnetic spring pushes against
the ``ceiling'' with an effective pressure $p_m \sim B_\varphi^2/8\pi$ due
to the toroidal field, $B_\varphi$.  As time goes on, more
toroidal loops form, and the toroidal field becomes stronger. 
(c)~At some later time, $t=t_2$, the pressure becomes so large that the
spring pushes away the ``ceiling''
and accelerates the plasma attached to it along the rotation axis,
forming a jet.  
Asymptotically far from the center, the toroidal field is the
dominant field component and determines the dynamics of the jet.
(d)~It is convenient to think of the jet as a collection of
toroidal field loops that slide down the poloidal field lines and
accelerate along the
jet under the action of their own pressure gradient and hoop stress. The
rotation of the sphere continuously twists the poloidal field into
new toroidal loops at a rate that, in steady state, balances the
rate at which the loops move downstream.
}
\end{figure}

How are jets magnetically launched?  Figure~\ref{jetcartoon} shows a
cartoon depiction of this.  Consider a vertical magnetic field line 
attached on one end to a perfectly conducting sphere, which represents the
central compact object, and on the other end to a perfectly conducting
``ceiling'' which represents the ambient medium (panel a).  As the sphere is
spinning, after $N$ turns the initially vertical field line develops
$N$ toroidal field loops (panel b).  This magnetic spring pushes against the
ceiling due to the pressure of the
toroidal field. As more toroidal loops form
and the toroidal field becomes stronger, the spring pushes away the ceiling
and accelerates any plasma attached to it along the rotation axis,
forming a jet  (panels (c) and (d) in
Figure~\ref{jetcartoon}, see the caption
for details). In the case when the central body is a black hole,
which does not have a surface, the rotation of space-time causes the
rotation of the field lines, and jets form in a similar fashion via a
process referred to as Blandford-Znajek mechanism (BZ, hereafter) \citep{bz77}.

\section{Radio-Loud/Quiet Dichotomy of Active Galactic Nuclei}
Despite decades of research, we still do not understand what determines
jet power in accretion systems or how to accurately infer jet powers
observationally \citep{rs91,ghi_blazars_2010,fernandes_agnjetefficiency_2010,mcnamara_agnjetefficiency_2010,punsly2011,mr11a}.  There are a number of unresolved
puzzles.  For the same nuclear B-band
luminosity, radio-loud (RL) AGN have a factor of $10^3$ higher total (core
plus extended) radio
luminosity than radio-quiet (RQ) AGN.  Figure~3 in \citep{ssl07} clearly
shows that these two flavors of AGN follow two well-separated tracks
on the radio-loudness diagram, which shows the radio-loudness
parameter, $\mathcal R$, the ratio of radio to optical luminosities at specified
frequencies vs. the Eddington ratio,
$\lambda$. This dichotomy is especially pronounced at low
$\lambda\lesssim0.01$ when the accretion systems are thought to be be in
a radiatively inefficient state of accretion, with a geometrically thick
accretion disk. For
higher $\lambda$-values, accretion systems can undergo spectral state
transitions (as seen in BHBs, \citep{fend04a}), which might cause 
differences in $\mathcal R$. Therefore, we limit ourselves to low-$\lambda$ systems. 

If total (core plus extended)
radio luminosity is a tracer of jet power, could the dichotomy be due
to the differences in the spin of central BHs that power the
relativistic jets in RL and RQ
AGN \citep{wc95}?  Characteristic values of BH spin in RQ
AGN, which lie predominantly at the centers of spiral galaxies, can be
quite low, $\abs{a}\lesssim0.3$. This can be understood if the
orientation of angular momentum accreted by the central BHs changes
randomly between accretion events \citep{ms96}. On the other hand, the spin of
central BHs in RL AGN, which are hosted predominantly by elliptical
galaxies, can be much higher, $a\sim1$ \citep{ms96}.  Can this
difference in BH spin lead to a factor of $10^3$ dichotomy?

\begin{figure}[t]
\begin{center}
\includegraphics[width=0.65\textwidth]{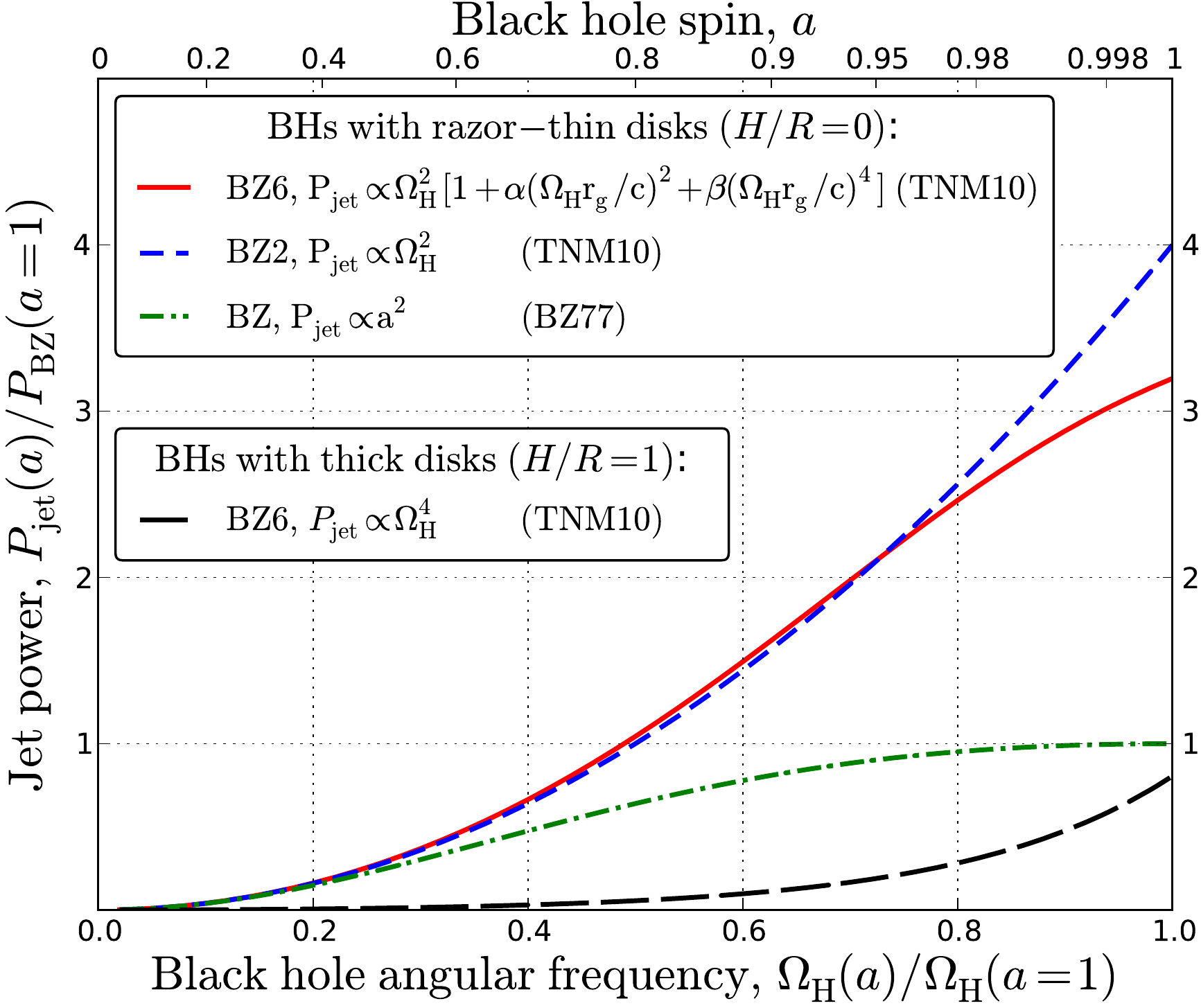}
\end{center}
\vspace{-0.5cm}
\caption{\label{figbz} Comparison of various approximations for
  jet power,  $P_{\rm jet}$, versus
  BH angular frequency, $\Omega_{\rm H}$ (lower $x$-axis), and BH
  spin, $a$ (upper $x$-axis).  All powers are normalized to the
  maximum achievable power in the standard low-spin BZ approximation, $P_{\rm
    jet}\propto a^2$, which is shown with dot-dashed green
  line.   The standard BZ approximation remains
  accurate only for moderate values of spin,
  $a\lesssim0.5$, and for maximally-spinning BHs  ($a=1$) it under-predicts
  the true jet power by a factor of $\approx 3$.  This can be seen by
  comparing to the BZ6 approximation, $P_{\rm jet}\propto
  \Omega_{\rm H}^2\left[1+\alpha (\Omega_{\rm H}r_g/c)^2+\beta (\Omega_{\rm
      H}r_g/c)^4\right]$, which is shown with red
  solid line and which is uniformly accurate for all values of $a$ (the
  constant factors, $\alpha\approx1.38$ and
  $\beta\approx-9.2$, are obtained in~\citep{tch10a}).  
  BZ2 approximation, $P_{\rm jet}\propto \Omega_{\rm
    H}^2$, shown with blue short dashed line, is accurate up to $a\lesssim0.95$,
  beyond which it requires a modest correction.
  If a BH is surrounded
  by a thick accretion disk, the disk intercepts BH power that is
  emitted into a meridional band occulted by the
  disk. 
  This reduces jet footprint and
  power relative to the full BH power for
  razor-thin disks given above and shown with red solid line 
  (see also \citep{tch10a}), provided that 
  BH magnetic flux is held constant.   
  The resulting BZ6 power for thick disks, with characteristic angular thickness
  $H/R=1$ expected in low luminosity AGN, 
  is shown with long dashed black curve.
  Disk occultation effect on jet power is most pronounced at low
  spins: it causes a substantial steepening of jet power dependence on
  spin and can explain a factor of $10^3$ radio loud/quiet dichotomy of AGN.
 }
\end{figure}

In the presence of large-scale magnetic fields, rotating BHs produce
outflows via the BZ mechanism,
which extracts BH rotational energy at a rate,
\begin{equation}
  \label{eq:bz}
  P_{\rm BZ} = \frac{\kappa}{4\pi c} \PhiBH^2 \frac{a^2}{16r_g^2}\qquad
  \textrm{(standard BZ formula, low-spin limit, } a^2\ll1 \textrm{)},
\end{equation}
where $\kappa\approx0.05$ weakly depends on magnetic field geometry,
$\PhiBH$ is the magnetic flux through the BH horizon, and $r_g=GM/c^2$
is BH gravitational radius \citep{bz77}. This
low-spin approximation, which we refer to as the standard BZ formula, 
remains accurate up to $a\lesssim0.5$
\citep{kom01,tn08}, which can be seen in Figure~\ref{figbz}.
Clearly, the $\propto a^2$ scaling does not appear steep enough to
explain the $10^3$ dichotomy: BH power varies by a factor of $\sim 10$ if $a$
varies from $0.3$ to~$1$. Can the power dependence becomes steeper as
$a\to1$?  Figure~\ref{figbz} illustrates that an expansion in the powers of BH angular
frequency, $\OmegaH=ac/2r_{\rm H}$, 
\begin{equation}
  \label{eq:bz6}
  P_{\rm BZ} = \frac{\kappa}{4\pi c} \PhiBH^2 \OmegaH^2\, f(\OmegaH)\qquad
  \textrm{ (BZ6 formula, accurate for all values of $a$)},
\end{equation}
remains accurate up to $a\lesssim0.95$ for $f=1$ (referred to as the
BZ2 formula), where $r_{\rm
  H}=r_g[1+(1-a^2)^{1/2}]$ is BH horizon radius. Equation~(\ref{eq:bz6})  with
$f(\OmegaH)\approx1+1.38  (\Omega_{\rm H}r_g/c)^2-9.2(\Omega_{\rm H}r_g/c)^4$
remains accurate for all spins \citep{tch10a}, and this gives us our 
BZ6 formula. Figure~\ref{figbz} shows that while at small spin the
BZ6 formula (\ref{eq:bz6}) and the standard BZ formula
(\ref{eq:bz}) agree, as $a\to1$ the BZ6 formula gives about $3$
times more power. However, the spin dependence 
is still not steep enough to explain the dichotomy: BH power varies by
a factor of $\sim30$ if $a$ varies from $0.3$ to $1$.

So far we considered razor-thin accretion disks, which is an implicit
assumption behind the above expressions for power, eqs.~(\ref{eq:bz})--(\ref{eq:bz6}).  However, disks 
in low-luminosity AGN are thick, with disk angular thickness
as large as $H/R\sim1$  (including disk body and magnetized corona).  
Such a thick disk mass-loads equatorial field lines on the BH, so that
they become part of a sub-relativistic disk wind, and 
only the magnetic flux in the highly magnetized funnel above and
below the disk contributes to the jet.  If the total magnetic flux, $\PhiBH$,
through the BH is held constant, this disk occultation effect can cause a much
steeper spin-dependence of jet power, $P_{\rm jet} \propto \OmegaH^4$,
as is seen in Fig.~\ref{figbz}.
Then, variation of power by a factor of $10^3$ is possible, which
can explain the radio loud/quiet dichotomy of AGN \citep{tch10a}.

\section{What Sets Jet Power in Black Hole Accretion Systems?}

\begin{figure}[t]
\begin{center}
\includegraphics[width=1\textwidth]{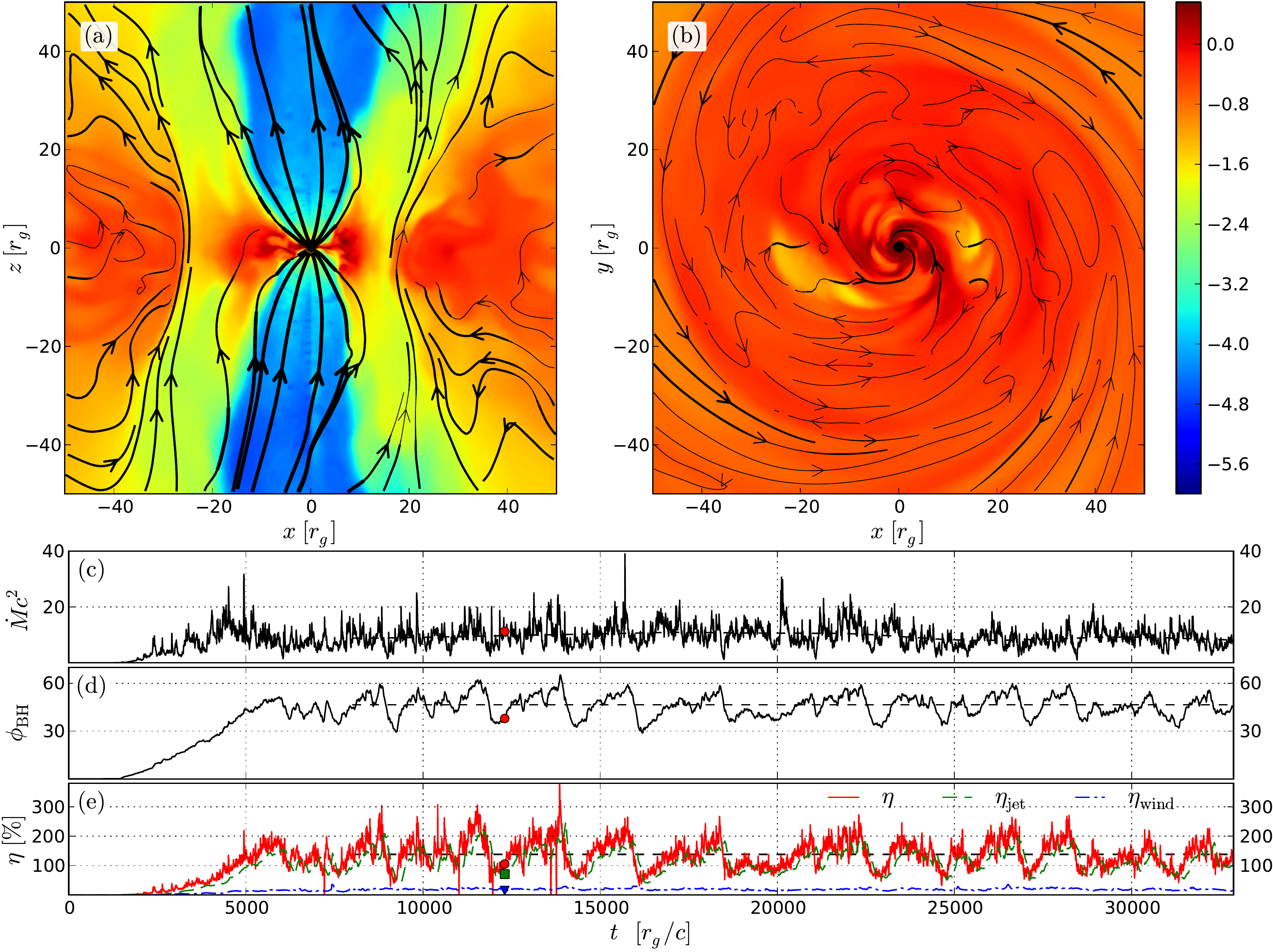}
\end{center}
\vspace{-0.5cm}
\caption{\label{movframe} A snapshot 
   and time-dependence of a magnetically-arrested
    accretion disk \citep{tch11}. A movie is available at
    \href{http://youtu.be/nRGCNaWST5Q}{http://youtu.be/nRGCNaWST5Q} . The
    top panels, (a) and (b), show the vertical ($x-z$) and horizontal ($x-y$) slices
    through the accretion flow. Color shows the logarithm of density:
    red shows high and blue shows low values (see color bar). The
    black circle shows a BH ($a=0.99$) and black lines show field
    lines. The bottom 3 panels show, from top to bottom, mass
    accretion rate, $\dot M$ (panel c), dimensionless flux through the BH, 
    $\phi_{\rm BH}$ (panel d), and outflow efficiencies of the whole
    outflow, $\eta$ (red solid line),
    the jet, $\etaj$ (green dashed line), and the wind, $\etaw$
    (blue dash-dotted line) (panel e). Since $\etaj$ and $\etaw$ are measured at
    $r=100r_g$, they lag by $\Delta t\approx100r_g/c$ relative to
    $\eta$, which is measured at $r=r_{\rm H}$.
     Colored symbols show values at the snapshot time, $t\approx12305r_g/c$. At $t=0$, the accretion flow contains a large amount of
    large-scale magnetic flux. 
    Accretion brings mass and flux to the
    hole, and $\phi_{\rm BH}$ increases until it reaches
    a maximum value at around $t\approx6000$ time units. At this time
    the BH is saturated with magnetic flux and produces as
    much power as possible. However, the accretion flow brings in even
    more flux, which impedes the accretion and leads to a
    magnetically arrested disk. Some of the flux
    escapes from the BH via magnetic interchange and flux eruptions, two of
    which are seen in panels (a) and (b), which frees up room for new
    flux. This process continues in a quasi-periodic fashion, and on
    average the BH produces outflows at $\eta\approx140$\%
    efficiency, i.e., outflows carry more energy than the entire
rest-mass energy supplied by accretion. 
 }
\end{figure}

We have shown that BH power is directly proportional to the square of
BH magnetic flux, $\PhiBH$, and the square of BH angular frequency,
$\OmegaH$ (see eq.~\ref{eq:bz6}), with small corrections beyond
$a\gtrsim0.95$.  In Nature,  $\PhiBH$ is a free parameter,
whose value is poorly observationally constrained.  
Clearly, we have $\PhiBH^2\propto \dot M c^2$.
But what sets the dimensionless ratio,
$\phibh^2= {\PhiBH^2}/{\MdotH r_g^2 c}$,
which characterizes the degree of inner disk
magnetization and controls energy extraction
from the BH
\citep{gam99,kb09,penna10,tch11,tm12a,mtb12}?
Using $\phibh$, we define BZ efficiency as BZ6 power
(eq.~\ref{eq:bz6}) normalized by $\dot Mc^2$:
\begin{equation}
  \label{eq:etabz}
  \etabz = \frac{\avg{P_{\rm BZ}}}{\avg{\MdotH}c^2} \times 100\%
         = \frac{\kappa}{4\pi} \avg{\phibh^2}\left(\frac{\OmegaH
             r_g}{c}\right)^2\, f(\OmegaH) \times 100\%,
\end{equation}
where $\avg{...}$ is a time-average. 
 Previous GRMHD simulations
found $\etabz\lesssim20\%$, even for nearly maximally spinning BHs
\citep{mck05,dev05a,hk06,bb11}. 
Are larger values of $\etabz$ possible?
This is an especially important question since observations suggest
that some AGN produce jets and winds at a high efficiency, $\eta=(P_{\rm jet}+P_{\rm
  wind})/\avg{\dot Mc^2}\times100\%\gtrsim100\%$
\citep{rs91,fernandes_agnjetefficiency_2010,ghi_blazars_2010,
punsly2011,mcnamara_agnjetefficiency_2010}, where $P_{\rm jet}$ and $P_{\rm
  wind}$ are jet and wind powers, respectively.
Are BHs capable of powering
such highly efficient outflows?
We tested this with global time-dependent GRMHD accretion disk-jet
simulations for different values of BH spin. As is
standard, we initialized the simulations with an equilibrium
hydrodynamic torus around a spinning BH and inserted a
weak, purely poloidal ($B_\varphi=0$) magnetic field loop into the torus. 
Clearly, jet efficiency (eq.~\ref{eq:etabz}) depends on
the magnetic flux, and time-dependent numerical
simulations show that the larger the large-scale
vertical magnetic flux in the initial torus, the more efficient 
the jets \citep{mck04,mck05,bhk07,mb09}. To maximize
jet efficiency, we
populated the torus with a much larger magnetic flux than in previous work.  In fact, our torus
contained more magnetic flux than the inner disk can
push into the BH.
The outcome for BH spin $a=0.99$ is shown in
Figure~\ref{movframe}.
The
magnetorotational instability (MRI, \citep{bal91}) brings
gas and magnetic flux to the BH, and an accretion disk of angular
density scale height, $h/r\approx0.3$, forms.  Figure~\ref{movframe}(d),(e)
shows that both $\phibh$ and $\eta$ increase until
$t\approx6000r_g/c$, beyond which they saturate and oscillate around
the mean, with time-average $\eta\approx140\%$. 
At this time, the BH is saturated with magnetic flux, and the
magnetic field on the hole is so strong
that it obstructs the accretion and leads to a magnetically-arrested
disk, MAD 
\citep{bkr74,bkr76,igu03,nia03,igu08,tch11,tm12a,mtb12}. 
In
this state both BH dimensionless
magnetic flux, $\phibh$, and outflow efficiency, $\eta$, are maximum, 
so it is not surprising that we find much higher values of
$\eta$ than previously reported \citep{tch11}.
In fact, $\eta>100\%$, therefore
jets and winds carry more energy than the entire rest-mass supplied by
the accretion, and this unambiguously shows that 
\emph{net} energy is extracted from the accreting BH. 
This is the first demonstration of \emph{net} energy
extraction from a BH in a realistic astrophysical scenario. 
Thicker MADs ($h/r\approx0.6$) produce
outflows at an even higher efficiency, $\eta\simeq300$\%~\citep{mtb12}.

\begin{figure}[t]
\begin{center}
\includegraphics[width=1\textwidth]{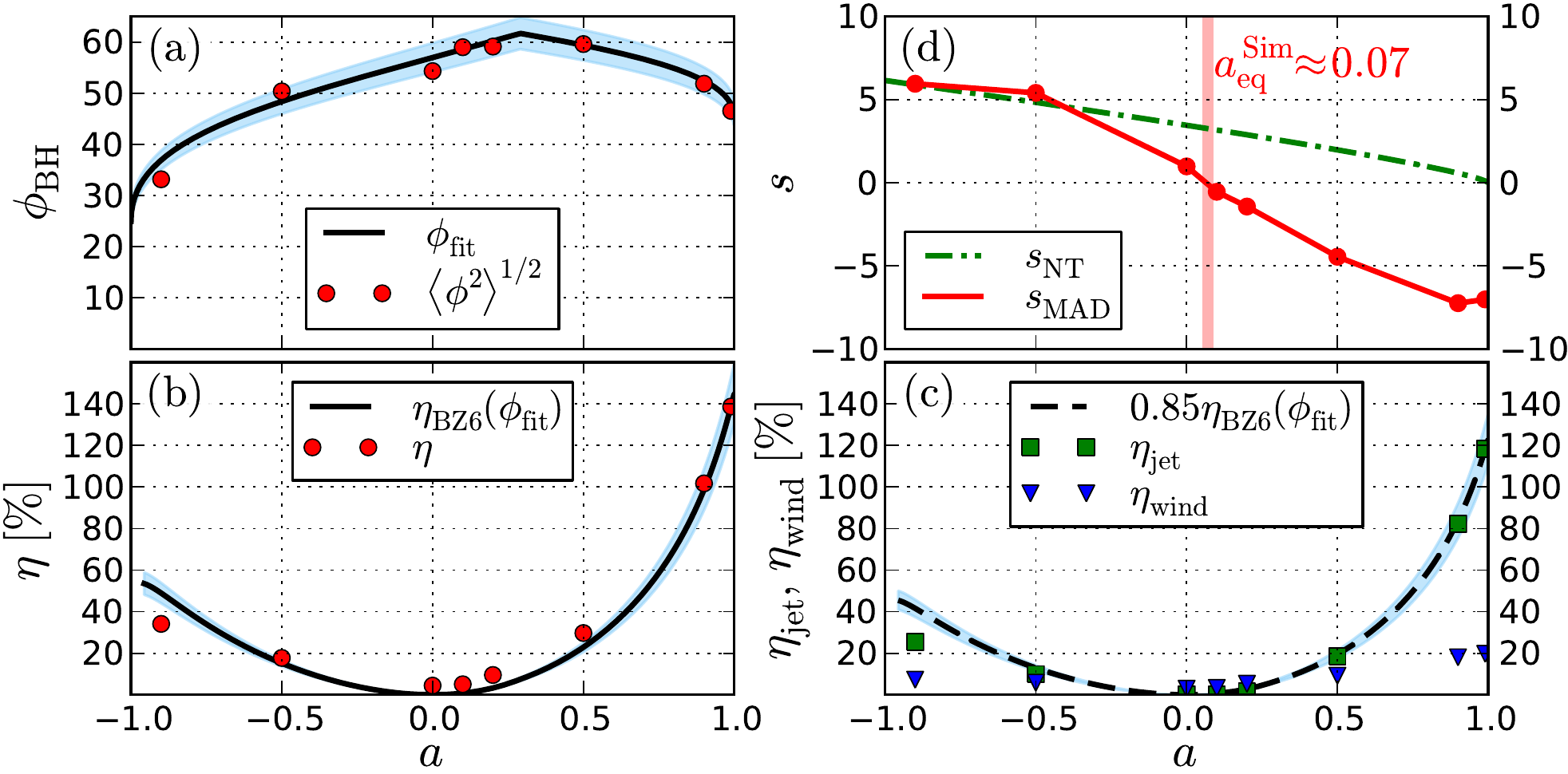}
\end{center}
\vspace{-0.5cm}
\caption{\label{jetwindeta} 
Spin-dependence of various quantities for MADs with $h/r\approx0.3$. (a) Dimensionless BH
spin, $\phibh$: simulation results (red dots), 
by-eye fit, $\phi_{\rm fit}$, comprised of two linear segments 
in a $\phibh$--$\OmegaH$ plane (black line), and 
$5$\% uncertainty on the fit (blue band). (b) Energy outflow
efficiency, $\eta$: simulation results (red dots), 
BZ6 efficiency (black line, eq.~\ref{eq:bz6}) assuming $\phibh(a)=\phi_{\rm
  fit}(a)$, and $10$\% uncertainty (blue band). (c)
Jet efficiency (green squares) and wind 
efficiency (inverted blue triangles). Dashed line shows $85$\% of the above BZ6 efficiency
(a good estimate of jet power for prograde BHs
\citep{tm12a}), and blue color shows a $10$\% uncertainty band. 
(d) BH spin-up parameter, $s$, for a
thin Novikov-Thorne disk \citep{nov73} (green dash-dotted line) and
for the simulations (red dots). Whereas for thin disks the equilibrium value of BH spin is
$a_{\rm eq}^{\rm NT}=1$, for our simulations it is much lower,
is $a_{\rm eq}^{\rm Sim}\approx0.07$  (vertical
red band). 
 }
\end{figure}

Importantly, in the MAD state $\eta$ is independent
of the initial amount of magnetic flux in the accretion flow, i.e.,
$\eta$ depends only on BH spin, $a$, and disk density angular thickness, $h/r$
\citep{tm12a}. 
This allows us to reliably study spin-dependence of various quantities, shown
in Figure~\ref{jetwindeta}. 
Dimensionless BH flux,
$\phibh$, shows $\sim50$\% variation with spin
(see panel~a). 
Prograde BHs ($a>0$) are
a few times more efficient than retrograde BHs ($a<0$) for the same value of
$\abs{a}$ (see panel~b).
Quite the opposite---$10$ times larger
efficiency of retrograde BHs than of prograde BHs---is predicted by
the ``gap'' model \citep{gar09}, which assumes that zero
magnetic flux populates the ``gap'' between the BH horizon and the innermost stable
circular orbit (ISCO). 
We find that this interpretation of the ISCO---as the surface inside of
which all magnetic flux nearly freely falls into the BH---does not
hold in MADs:
interchange instability \citep{ss01} and flux eruptions (see
Fig.~\ref{movframe})
populate the ISCO interior with flux.
In fact, the thinner the disk, the larger the fraction of BH magnetic
flux that
resides in the region between the BH horizon and the ISCO 
\citep{mtb12,tm12a}. 

Panel (c) shows the division of total outflow
efficiency into highly magnetized jet and weakly magnetized
wind components, with efficiencies $\etaj$ and $\etaw$,
respectively. Since jets are BH spin-powered 
(eq.~\ref{eq:bz6}), for $a=0$ jet efficiency vanishes,
but winds still derive their power 
from an accretion disk via a Blandford-Payne--type mechanism \citep{bp82}. The
larger the spin, the more efficient jets and winds. 
For rapidly spinning BHs most of the energy---about
$85$\% for prograde BHs---is carried by
relativistic jets.  Importantly, even rather slowly spinning BHs, with
$a\lesssim0.5$, produce prominent BH spin-powered jets.
This is in agreement with
the recent evidence that jets in BHBs are powered by 
BH rotation over a wide range of BH spin, $0.1\lesssim a\lesssim0.9$ \citep{nm12}.

Do our highly efficient jets affect the spin of central BHs?  
Figure~\ref{jetwindeta}(d) shows spin-dependence of BH spin-up
parameter, $s=M/\dot M \times \textrm da/\textrm dt$
\citep{gammie_bh_spin_evolution_2004}. 
For standard geometrically thin
accretion disks, $s>0$ (see
Figure~\ref{jetwindeta}(d) and \citep{nov73}), and the equilibrium
spin is $a^{\rm NT}_{\rm eq}=1$
\citep{bardeen70} (we neglect photon capture by the BH, which would
limit the spin to $a\approx0.998$ \citep{thorne74}). Thick
accretion flows in time-dependent numerical simulations
\citep{mck04,gammie_bh_spin_evolution_2004,kro05} and
semi-analytic studies \citep{benson09} typically have $a_{\rm
  eq}\sim0.9$ (however, see \citep{msl98}). 
Figure~\ref{jetwindeta}(d) shows that 
the equilibrium value of spin for our MADs (with
$h/r\approx0.3$) is much smaller, $a_{\rm eq}^{\rm
  Sim}\approx0.07$, due to large BH spin-down torques by our powerful jets. Since jet efficiency is  
$\sim10^3$ times lower at this value of spin than for rapidly spinning BHs, such equilibrium spin
systems are possible candidates for RQ AGN.

\section{Conclusions}
We confirm that the standard BZ power formula, $P_{\rm jet}\propto
a^2$ (eq.~\ref{eq:bz}), remains accurate only for $a\lesssim0.5$ and present a higher order
 BZ6 approximation (eq.~\ref{eq:bz6}) that is accurate for all values of spin
(see Figure~\ref{figbz}).
We find that if the BH magnetic flux is held constant,
the presence of a thick accretion disk
in low-luminosity AGN leads to a steep dependence of
BH jet power on spin, $P_{\rm jet}\propto\OmegaH^4$. This steep
dependence allows to explain a factor of $10^3$ radio
loud/quiet dichotomy of AGN via having two galaxy populations
different by the spin of central BHs.

We carried out a series of time-dependent global GRMHD simulations of
BH accretion that contain a large amount of magnetic flux.   We show that the accumulation of
magnetic flux around the center can lead to the formation of a strong
centrally-concentrated magnetic field that saturates the BH, obstructs the accretion, and
leads to a magnetically-arrested disk (MAD,
\citep{bkr74,bkr76,igu03,nia03,igu08,tch11,tm12a,mtb12}). 
We show that in this state the outflow efficiency, $\eta$, depends only on the BH
spin, $a$, and the angular density thickness of the accretion flow,
$h/r$, and is independent of the initial amount of magnetic flux in
the disk.  Since BH
magnetic flux is as large as possible, we expect MAD systems to
achieve the
maximum $\eta$ for a given BH spin and disk
thickness. Indeed, we find highly efficient outflows, with
$\eta\gtrsim100\%$, which suggests that MADs 
might explain observations of AGN with apparent $\eta\sim
\mathrm{few}\times100\%$ \citep{rs91,ghi_blazars_2010,fernandes_agnjetefficiency_2010,mcnamara_agnjetefficiency_2010,punsly2011,mr11a}.
We determine the spin-dependence of jet, wind, and total
efficiencies in our simulations (Figure~\ref{jetwindeta}) and expect that this
information can be useful for calibration of semi-analytic jet power
models, for studies seeking to infer BH spin from the observed jet
efficiency, and for tests of general relativity \citep{daly2011,mr11a,gne11,lz11,bam12}.
\vspace{0.15cm}
\newline
AT was supported by the Princeton Center for
Theoretical Science fellowship. AT and RN were supported in part by
NSF grant AST-1041590 and NASA grant NNX11AE16G. We acknowledge
NSF support via TeraGrid resources:
NICS Kraken and Nautilus, where simulations were carried out
and data were analyzed, and NCSA MSS and TACC Ranch, where data
were backed up, under grant numbers
TG-AST100040 (AT), 
 TG-AST080025N (JCM),  TG-AST080026N (RN).

\providecommand{\newblock}{}

\end{document}